# Broadband high-resolution integrated spectrometer architecture & data processing method

Mehedi Hasan[1,2*], Gazi Mahamud Hasan[1], Houman Ghorbani[1], Mohammad Rad[2], Peng Liu[1], Eric Bernier[2] and Trevor Hall[1]

[1]Photonic Technology Laboratory, Centre for Research in Photonics, Advanced Research Complex, University of Ottawa, 25 Templeton Street, Ottawa, K1N 6N5, ON, Canada
[2]Huawei Technologies Canada, 303 Terry Fox Drive, Kanata, K2K 3J1, ON, Canada
*mhasa067@uottawa.ca

**Abstract:** Up-to-date network telemetry is the key enabler for resource optimization by capacity scaling, fault recovery, and network reconfiguration among other means. Reliable optical performance monitoring in general and, specifically, the monitoring of the spectral profile of WDM signals in fixed- and flex- grid architectures across the entire C-band, remains challenging. This article describes a two-stage spectrometer architecture amenable to integration on a single chip that can measure quantitatively the spectrum across the entire C-band with a resolution of ∼ 1 GHz. The first stage consists of a ring resonator with intra-ring phase shifter to provide a tuneable fine filter. The second stage makes use of an AWG subsystem and novel processing algorithm to synthesize a tuneable coarse filter with a flat passband which isolates individual resonances of a multiplicity of ring resonances. Due to its maturity and low loss, CMOS compatible $Si_3N_4$ is chosen for integration. A fabricated ring resonator functioning over the entire C-band with 1.3 GHz FWHM bandwidth resonances tunable over a complete free spectral range of 50 GHz is experimentally demonstrated. The complete system operation is demonstrated using an industry standard simulation tool and AWG constructor data. The operation of the circuit is invariant to the optical path length between individual components substantially improving robustness to fabrication process variations.

**Introduction**
In an optical network, optical performance monitoring (OPM) is the key enabling technology for reliable spectrum management. Up-to-date network telemetry is required for capacity scaling, network or component fault recovery, and network reconfiguration through performance prediction and planning. The monitoring includes information on bit-error-rate (BER), optical signal-to-noise ratio (OSNR), electrical signal-to-noise ratio (ESNR), loss, and power. The monitoring information is then provided to the network management agent for resource optimisation to maximise the reach versus rate. In practice, optical performance monitoring may be based on the measurement of one or several parameters. However, OPM is used herein to refer to "power" monitoring since power is one of the key indicators of performance in optical systems.

In transport optics, especially in WDM networks, spectral sensing is not straightforward. Traditionally, WDM channels are distributed over the 40 nm wide fibre-optic C-band (1530 nm to 1565 nm) with fixed centre frequencies arranged on the International Telecommunication Union (ITU) grid at intervals of 50 GHz or 100 GHz and an OPM card is used to measure the power. The working principle of the OPM card involves sweeping a tunable filter with 50 GHz resolution over the spectrum to make available ITU grid channel power readings. Due to their excessive power consumption, size, and cost, OPM cards are deployed only at a few points in the network; typically co-located with reconfigurable add-drop modules (ROADMs). However, current optical networks are elastic in nature, i.e., the channels are not located on a fixed regular grid, rather the channel centre wavelength can be placed at an arbitrary location within the spectrum. The flexible grid can support a variety of channel power profiles (i.e., bandwidth and power spectral density) with a channel centre frequency placement resolution as fine as 6.25 GHz. Flex-ready spectrum measurements are required to facilitate the deployment of the flex-grid system. As a result, flex-grid ready ROADM architectures are equipped with new modules that can measure power at a desired frequency location and resolution. However, due to cost issues, spectrum measurement is only performed at add-drop nodes and not at amplifier nodes. Moreover, a single OPM module is shared by the multi degree-ROADM, so the OPM measurement speed reduces as the number of lines it supports increases. Consequently, the performance of WDM channels in a section (ROADM-to-ROADM) is modeled based on an analytical or a semi-analytical analysis or a machine learning approach. The absence of OPM makes it difficult to have live and accurate network measurement; and hence hard to implement performance optimization. Complete knowledge of spectral content in a network is a prerequisite for the effective use of color-, direction-, contention-, grid-, filter-, gap- less ROADM; flexible modulation formats; flexible channels frequencies; and spectral assignment.

A variety of different approaches to the problem of spectral sensing with high resolution across a wide band have been disclosed in the literature [1-9] but when scaled to combine acceptable resolution with wideband operation their practical implementation is most often not feasible due to excessive cost, loss, and footprint. An integrated solution for a high resolution (sub-GHz) spectrometer to monitor the power in fixed- and flex- grid architectures across the entire C band 1530 nm to 1565 nm remains challenging. Nevertheless, a three-stage architecture proposed in a previous publication [10] has been shown to be viable. The first stage is a tunable ring resonator (RR) that defines the resolution. The third stage is an arrayed waveguide grating (AWG) that isolates one RR resonance within each of its channels. The principal innovation is the ganged tuning of the RR and AWG to retain the RR resonance at the center of the AWG channel passband. This is achieved by a second stage that uses an MZI with delay imbalanced arms to form a coherent superposition of two interleaved AWG channel spectra corresponding to a pair of input ports. Further details can be found in [10]. This article describes a refinement of the architecture and method that enables the ganged tuning of the AWG to be virtualized. The coherent superposition of a pair of AWG input channels is replaced by the incoherent superposition of pairs of AWG outputs, which may be performed by processing the measured AWG channel powers. The MZI stage is eliminated, releasing the spectrometer from any requirement to control inter-stage optical path lengths, and thereby significantly easing manufacture.

In this paper, the circuit architecture design, modelling and integration feasibility of an ultra high-resolution wideband spectrometer based on the refined architecture is presented. The purpose of the proposed circuit is to measure the spectral profile of WDM channels in flex- and fixed-grid architectures across 1530 nm to 1565 nm (C band). The architecture combines a RR and AWG subcircuit with a simple but novel information processing method that enables the spectrometer to scan the entire C-band with high resolution (~1GHz) using only one dynamic control. A three AWG configuration offers substantially zero adjacent channel leakage but for clarity of exposition this article focusses on the two AWG minimum configuration. Hardware economies may be made by replacing multiple AWG by a single time multiplexed multiple-input AWG. Details of the three AWG configuration may be found in reference [11].

The proposed spectrometer can be fabricated using any suitable photonic integrated circuit fabrication platform. However, the resolution bandwidth of the spectrometer primarily depends on the RR waveguide loss (dB/turn). Hence to meet the specification of the proposed spectrometer, the CMOS compatible $Si_3N_4$ photonic integration platform has been selected as it offers low loss, tight confinement, low dispersion, and a mature thermo-optic phase shifter technology. There are ample reports in the literature of the technological verification of all components required by the spectrometer fabricated using the $Si_3N_4$ integration platform [12-14], including a high port count AWG [15]. The detail simulation verification of the proposed architecture is presented using fabricated components (RR) and components (AWG) designed for fabrication. A combination of industry standard software tools is used for the component design. Owing to the information processing method, the operation of the circuit is invariant to the optical path length between individual components substantially improving robustness to fabrication process variations.

**Operating principle**
The spectrometer architecture consists of a first stage fine filter followed by a second stage coarse filter. The fine filter is implemented using a ring resonator (RR) which generates a comb of regularly spaced narrow bandwidth resonances. The RR frequency response is periodic with period known as the free spectral range (FSR). The bandwidth of an individual RR resonance defines the resolution which should be $< 1\ GHz$ to meet the target specification. An intra-ring phase shifter is used to tune the RR so that the comb of resonances is translated in frequency over one FSR. This ensures that spectral information is collected over a continuous frequency interval spanning the whole C-band.

The task of the coarse filter is to isolate an individual RR resonance among the multitude of resonances of the comb. This task requires the coarse filter to substantially extinguish all adjacent RR resonances while substantially transmitting the individual RR resonance irrespective of the frequency to which it is tuned. Since the comb of RR resonances is translated in frequency by the tuning mechanism over an interval $f \in (-\Delta f, \Delta f]$ spanning one FSR, to satisfy both these requirements, a fixed frequency ideal filter must have unit transmission over $(-\Delta f, \Delta f]$ and zero transmission over the complement of this interval. A practical fixed frequency filter cannot meet this requirement and would not only introduce undesirable attenuation within the passband but also introduce unacceptably high adjacent channel leakage at the limits of the tuning range. It is necessary that the coarse filter must be tuneable so that it can maintain the wanted resonance within a flat region of its passband and unwanted adjacent resonances within its stop band.

The analysis of virtual channel synthesis that follows is applicable to pre-detection (optical) processing, in which case all quantities are complex, and post-detection (electronic) processing, in which case all quantities are real and positive. The second stage of the architecture synthesizes a tuneable coarse filter channel by forming a weighted sum $y$ with weights $w_j$ of the outputs $x_j$ of a collection of physical channels indexed by $j$ with fixed centre frequency:

$$y = \sum_j w_j^* x_j \qquad (1)$$

The physical channels are provided by one or more arrayed waveguide grating (AWG) devices which together offer a channel frequency spacing that is equal to the RR FSR divided by an integer $N \geq 2$. Collectively, the passbands of the physical channels cover the entire C-band so that each RR resonance falls within the passband of at least one physical channel. Moreover, if a physical channel has a RR resonance falling within its passband, all other RR resonances fall within the stop band of that channel. The values of the weights are chosen to meet the two requirements specified in the preceding and consequently are a function of the tuning frequency. Following similar reasoning to maximal ratio combining [16], the weights are set equal to the complex conjugate of the transmission function of their associated physical channel which results in a synthesised channel transmission function:

$$H(\omega; \omega_0) = \frac{\sum_j w_j^*(\omega_0) w_j(\omega)}{\sum_j w_j^*(\omega_0) w_j(\omega_0)} \quad (2)$$

where the normalisation ensures $H(\omega_0; \omega_0) = 1$ satisfying the transmission requirement. A passband flattened AWG design is consequently not required. The design freedom thereby released assists the reduction of spurious AWG channel sidelobes and hence the attainment of the extinction requirement.

If the physical channel transmission falls rapidly outside its passband, only the pair of overlapping channels with centre frequencies adjacent to the RR resonance frequency contribute significantly to the weighted sum. The worst-case adjacent channel leakage occurs when the RR resonances frequency is midway between adjacent physical channel passband centre frequencies, i.e., the adjacent ring resonance is detuned from the passband centre of the contributing physical channels by a multiple $2N - 1$ of the passband half-width. For a Gaussian channel profile and a passband width specified by the FWHM then the adjacent channel leakage is suppressed by a factor of $2^{(2N-1)^2}$ or $27\ dB$ for $N = 2$ and $75\ dB$ for $N = 3$. The adjacent channel leakage consequently improves rapidly with $N$. However, the number of photodetectors required also scales with $N$. Consequently, the focus of this paper is on the $N = 2$ case. The adjacent channel leakage for $N = 2$ can be improved by slight reduction in the $-3dB$ bandwidth of the physical channels.

In prior work [10], a method was proposed for forming the weighted superposition of the fields generated within an AWG by a pair of input access waveguides driven by a delay imbalanced Mach-Zehnder interferometers (MZI). Since the processing occurs in the optical domain the tuneable coarse filter may directly drive further optical stages which may be advantageous in some applications. The coherent summation however requires tight control of the phase of the field and hence close integration of the MZI, and AWG is necessary to ensure optical path lengths are matched. The requirement to control the phase also introduces the requirement to switch between two pairs of input waveguides each driven by its own MZI (see reference [10] for details).

In the case of the spectrometer application may be performed after photodetection. All variables and functions in the weighted sum are then real valued and positive. This eliminates any requirement to control path lengths exterior to the RR and AWG components. The complete architecture is thereby rendered robust to fabrication process variations. It is even possible to use optical fibre to interconnect RR and AWG components implemented in different material platforms best suited to the component such as $Si_3N_4$ for the RR and doped silica for the AWG, albeit a goal of this work is single chip implementation.

There is some flexibility in the choice of the dependence of the weights as a function of tuning. The weight functions only need to approximate their associated channel profiles accurately over their passband and may rapidly and smoothly decay towards zero outside their passband. This expedient to some extent suppresses the sidelobe structure of practical AWG channel profiles. However, it cannot suppress crosstalk introduced by the main contributing channel, i.e., a low level of spurious lobes within one RR FSR from the passband of the channel main lobe is necessary to avoid adjacent channel leakage.

One embodiment of the proposed $N = 2$ AWG architecture is shown schematically in Fig. 1(a) as an example. A RR first stage provides the fine filter while the second stage AWG pair provide the coarse filtering. The RR generates a train of narrow resonances spaced in frequency by its free spectral range (FSR), while the AWG pair isolates each RR resonance in one or other or both of their output channels. Hence, the output channel frequency spacing of each AWG must be equal to the RR FSR by design. The RR defines the spectrometer resolution bandwidth and is tunable in frequency over one FSR. A nominally identical AWG pair with $-3\ dB$ channel passband-width close to half their output channel frequency spacing are required for this example. However, the AWG pair are not driven by the same input channel (port), rather they are driven by adjacent input ports as shown in Fig. 1(a). The input channel frequency spacing is equal to half of the output channel frequency spacing (i.e., $1/2$ FSR). The channel spectra of $AWG_1$ and $AWG_2$ thereby are interlaced and overlap as illustrated in Fig.1(b) for a 50 GHz channel frequency spacing. For clarity, only three of the channel spectra of each AWG are shown with possible mappings of the comb of RR resonances. In practice, 50 GHz AWGs require 88 channels to cover the whole C-band.

The comb of RR resonances is tuned by an intra-ring phase shift $\theta$. The translation in frequency of the comb is proportional to the phase shift and ranges over one FSR as $\theta$ ranges over $2\pi$ radians. For simplicity of exposition, $\theta = 0$ is taken to correspond to the alignment of the RR resonances with $AWG_1$ channel passband centre frequencies. It follows that $\theta = \pi$ corresponds to the alignment of the RR resonances with $AWG_2$ channel passband centre frequencies and $\theta = \pi/2, 3\pi/2$ corresponds to alignment respectively with the intersection between the upper (lower) $AWG_1$ -3dB channel passband edge and the lower (upper) $AWG_2$ -3dB channel passband edge. The optical power of each output channel of both AWGs is measured by a photodetector array while the RR is scanned over an FSR under the control of the data acquisition system (DAQ). The principal novelty of the spectrometer is the construction of a virtual AWG that is tuned to retain the ring resonance within the passband of its synthesised channels. A virtual channel with index $m$ is synthesised by the weighted sum of the optical power of the interlaced AWG$_1$. The weights depend on the frequency to which the RR is tuned or equivalently the tuning phase. For simplicity of exposition a Gaussian weighting function with the same FWHM as the AWG channel is convenient.

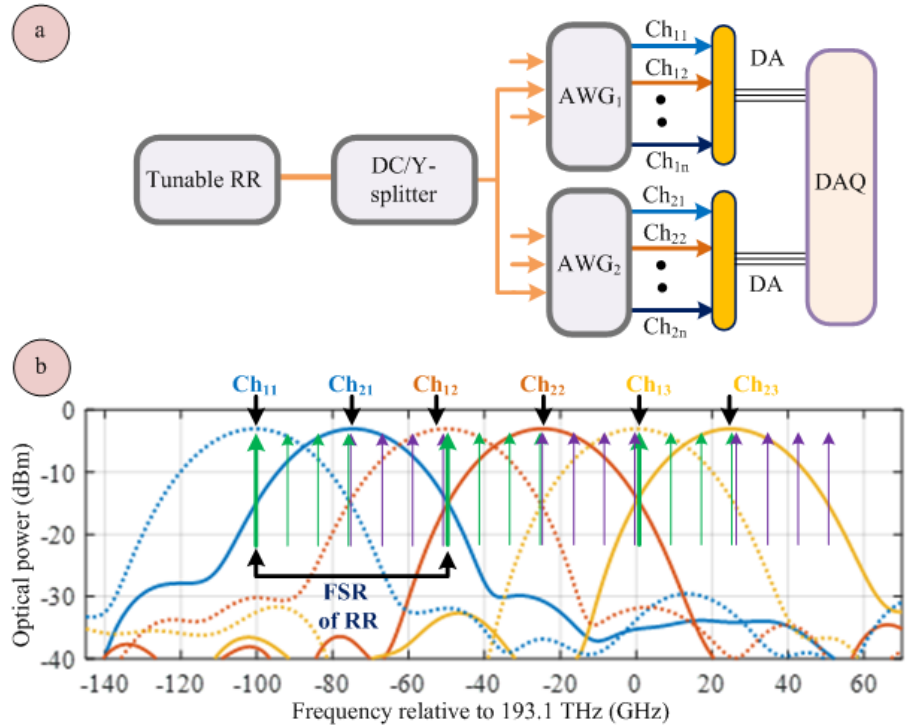

Fig. 1. (a) Schematic of the proposed spectrometer; (b) interlaced optical spectrum of $AWG_1$ and $AWG_2$ with the resonance mapping over one FSR. AWG, arrayed waveguide grating; RR, ring resonator; DC, directional coupler; DA, detector array; DAQ, data acquisition Ch; channel; FSR, free spectral range.

As the ring resonance is tuned over the first half of the FSR by a tuning phase shift from $\theta = 0$ to $\theta = \pi$ (green stem in Fig. 1(b)), AWG$_1$ channel $m$ (Ch$_{1m}$) and AWG$_2$ channel $m$ (Ch$_{2m}$) make the most significant contributions to the weighted sum. The contribution by Ch$_{1m}$ (Ch$_{2m}$) is weighted essentially by unity (zero) at $\theta = 0$ and by essentially zero (unity) at $\theta = \pi$. At $\theta = \pi/2$ the contributions of Ch$_{1m}$ and Ch$_{2m}$ are equal, and each is weighted by one half. Similarly, as the ring resonance is tuned over the second half of the FSR by a tuning phase shift from $\theta = \pi$ to $\theta = 2\pi$ (purple stem in Fig. 1(b)), AWG$_2$ channel $m$ (Ch$_{2m}$) and AWG$_1$ channel $m+1$ (Ch$_{1(m+1)}$) make the most significant contributions to the weighted sum. The contribution by Ch$_{2m}$ (Ch$_{1(m+1)}$) is weighted essentially by unity (zero) at $\theta = \pi$ and by essentially zero (unity) at $\theta = 2\pi$. At $\theta = 3\pi/2$ the contributions of Ch$_{2m}$ and Ch$_{1(m+1)}$ are equal, and each is weighted by one half.

The normalisation of the weights ensures that the synthesised channels have a perfectly flat passband to the extent that the weighting function closely approximates the channel profile over its passband. The error in the approximation creates some ripple. However, the ripple is small as the synthesised channel spectral profile is in close agreement with the ideal when the ring resonance is aligned at either passband centre or at the intersection of the -3dB passband edges of the AWG channels summed.

**Table 1. Detail design specifications of the proposed circuit architecture shown in Fig. 1(a)**

| Design specifications | Numbers | Remarks |
|---|---|---|
| RR free spectral range (GHz) | 50 | |
| AWG output channel spacing (GHz) | 50 | equal to the free spectral range of RR |
| AWG output channel bandwidth (GHz) | 20 | $\sim \leq 1/2$ AWG output channel spacing |
| AWG input channel spacing (GHz) | 25 | half of AWG output channel spacing |
| Number of AWG output channel | 88 | |
| Total spectrum covered (THz) | 4.4 | $88 \times 50 = 4400$ GHz (entire C band) |

The data processing is performed in the electronic domain by the data acquisition system which also controls the RR tuning phase shifter. The construction of the weights requires knowledge of the RR resonance frequency or a stable calibration of the phase shifter characteristics. Precision tracking of the position of a resonance is not mandatory for successful virtual channel synthesis given the wide AWG passband relative to the RR bandwidth. Nevertheless, for an integrated wavelength meter concept presented in [17] can be used to monitor the position of the RR resonance within a single channel of the AWG. Since the resonances are substantially periodic in frequency, albeit very slightly detuned by chromatic dispersion, the position of the resonance within each channel of the AWG can be determined easily.

The spectrometer requires only one control, which sets the intra-ring phase shift to tune the RR resonant frequency comb cyclically over one FSR. Device and circuit simulations; previously reported experimental demonstrations; and the process development kit support the practicality of a 50 GHz FSR RR. Table 1 shows the detail specifications of the proposed circuit design. For a RR FSR of 50 GHz, the required number of AWG output ports is 88 to cover the entire C-band. The number of output ports can be reduced by increasing the FSR of the RR, since 50→150 GHz channel spacing AWG having output ports up to 96 are available commercially [18]. On the other hand, a RR having FSR of 220 GHz fabricated using double strip TrIPleX$^{TM}$ waveguide technology is already reported in [3]. Hence, the proposed number of output channels of AWG can be reduced by a factor of 3 (~30 channels).

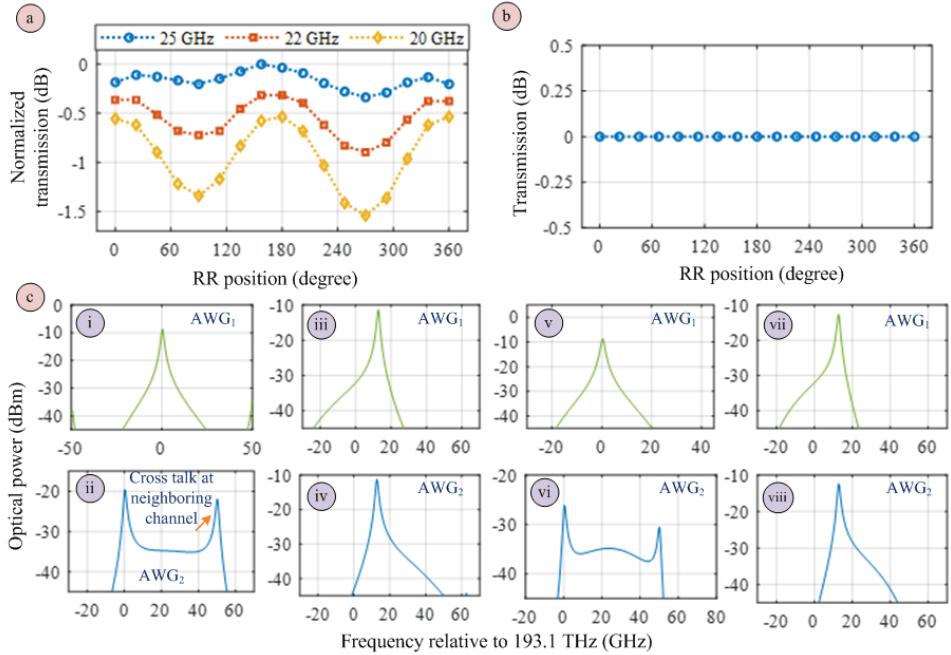

Fig. 2. Simulated optical power transmission as a function of ring resonance for different passband width of the AWG channel, (a) without weight summation, (b) with weight summation; (c) optical spectrum of $AWG_1$ and $AWG_2$ at channel one for ring resonance position of 0° and 90° for the passband width of 25 GHz [i-iv] and 20 GHz [v-viii] respectively.

The VPIphotonics simulation tool is used to evaluate the performance of the combined circuit architecture and data processing method. The optical power at each output channel of both AWGs is monitored while scanning the RR over one FSR. The calculation of the power depends on the number of samples used in the simulation time window. Since the signals were widely spread, a moderate time window is used in the simulation to avoid an excessive memory requirement. A slight variation in the result is obtained for different time window settings. The simulated output is then processed according to the described the algorithm. Figure 2(a) shows the variation in the measured total optical power (arithmetic sum of $AWG_1$ and $AWG_2$) as a function of the resonance position over one FSR for various AWG channel passband widths. Channel 1 & 2 of $AWG_1$ and channel 1 of $AWG_2$ are used for the calculation. The results show that the spectral measurement is almost flat with little ripple (~0 dB) over the entire FSR for a passband width of 25 GHz. Figure 2(c[i-iv]) shows the optical spectrum of $AWG_1$ and $AWG_2$ at output channel one (Ch11 & Ch21) for ring resonance position of 0° and 90° respectively for a passband width of 25 GHz. It shows that a strong (~-21→-22 dB) adjacent ring resonance component is present in the $AWG_2$ channel. Figure 2(c[v-viii]) shows the optical spectrum of $AWG_1$ and $AWG_2$ at the same resonance position for the pass band width of 20 GHz. The adjacent resonance cross talk is reduced to $\sim -30$ dB while maintaining minimal ripple. Figure 2(b) shows the transmission spectra using weighted summing method. As described earlier, this method can eliminate the ripple found in Fig. 2(a) along with improved cross talk. Since the transmission characteristic of an AWG is substantially periodic, the simulated result will almost be same for any other channel measurement of the AWG. The entire channel spectra of both AWGs are presented in Fig. 6. To substantially eliminate adjacent channel leakage, the architecture can further be upgraded to a $N = 3$ AWG configuration. The details can be found in reference [11].

**Validation & integration feasibility**
The feasibility of the ring resonator is verified experimentally while the system is demonstrated using an industry standard simulation tool. A combination of software simulation tools is used to validate the concept. VPIphotonics is used for the circuit simulation while Photon Design is used to verify the function of each component. The proposed spectrometer can be fabricated in any mature low-loss photonic integration platform with sufficient index contrast to support ring resonators. If the excess loss of the ring per turn is negligible in comparison to the power coupled out per turn, the resolution bandwidth of the spectrometer is determined by the power cross-coupling ratio of the couplers. The ring excess loss per turn consequently limits the achievable resolution. An integration platform supporting the design of low-loss waveguides and low-loss waveguide bends is therefore paramount. Owing to its low loss, tight confinement, low dispersion waveguides and a mature thermo-optic phase shifter technology, the CMOS compatible $Si_3N_4$ photonic integration platform offered by LioniX International was selected to meet the specification of the proposed spectrometer circuit. The platform also offers good prospects for further loss-reduction [19] and to lower power consumption, temperature insensitive, alternatives to thermo-optic phase-shift elements [20-21].

For effective use of resources, the fabrication plan envisaged multi-project wafer runs (MPW) for test structures followed by a custom wafer fabrication run for prototypes for demonstration. The LioniX MPW runs supports designs using the asymmetric double strip (ADS) waveguide only and the low-cost photolithography used has a minimum feature size of 1 $µ$m. Accordingly, the components and sub-circuits that constitute the proposed spectrometer are designed and simulated using ADS as reference waveguide. The waveguide characteristic over full C-band is obtained by using the Photon Design software tool FIMMWAVE. The TE-like mode is used in all the simulation due to its tight confinement, hence it exhibits lower bend loss in comparison to the TM-like mode. The effective group index of the mode at the at the smaller wavelength edge (1530 nm), centre wavelength (1545 nm), and the longer wavelength edge (1565 nm) of the C-band is found to be 1.7725, 1.76841 and 1.7629 respectively.

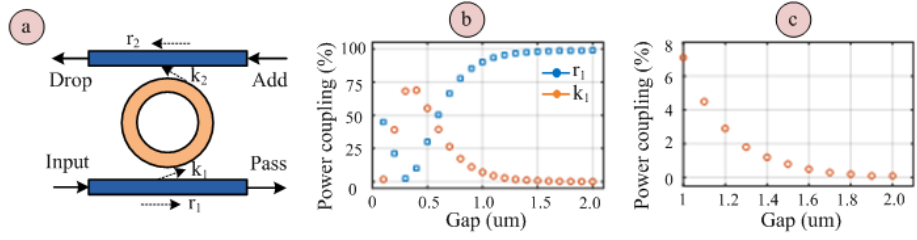

Fig. 3. (a) Schematic diagram of a RR; (b) power coupling between the ports of a directional coupler as a function of spatial separation among them; (c) zoom-in view of the power coupling at the cross port as a function of spatial separation.

Figure 3(a) shows the schematic diagram of a RR. As shown, it requires two directional couplers (DC). The resolution of the spectrometer is largely set by DC power cross coupling and is determined by the spatial gap between the interacting waveguides. A variety of numerical methods and a custom quasi-analytical method involving different approximations are used to bracket a range of gaps targeting 2-6% power coupling. Ring resonators with FSR in the required range are a challenging to simulate using standard tools. 3D FDTD requires too large computational resources. 2.5D FDTD uses effective index mode (EIM) solvers which cannot correctly model couplers. The eigen mode expansion (EME) method can be used in conjunction with a circuit simulator (PICWAVE, VPI) for very small FSR large racetrack rings, but it displayed an excessive computational power loss in the curved waveguide couplers of the 50 GHz FSR ring. The custom method integrated the effective index difference of the two fundamental symmetric and antisymmetric local eigenmodes of the coupled region found using MOLAB (any mode solver could be used) using an analytic adiabatic model that predicts the overall power transfer matrix of the proximate curved waveguides. An OptiBMP simulation, although strictly outside its domain of validity, provides a scan of the power coupling as shown in Fig.3 (b) & (c) similar to the quasi-analytic result. The range of gaps predicted is then sampled by test structures to enable the gap to be refined experimentally. The zoom in view of the power coupling at the cross port of the DC is shown in Fig. 3(c). The simulation results show that, for a gap of 1.3 um the power coupling at the cross port is 5% and it reduces to <~1% for a gap of 1.7 um. For the MPW run, several ring resonators are laid out on the mask having gaps from 1.2 um to 1.8 um with 0.2 um increment with the objective that at least one RR works well.

Two identical DC are used in the ring design. The ADS waveguide ring circumference is calculated to be 3.3928 $mm$ at 1545 nm for an FSR of 50 GHz using the following formula.

$$FSR(\omega) = c/n_g(\omega)l \qquad (3)$$

where, $n_g$ is the group index, $l$ is the length of the delay line to obtain the specified FSR. The corresponding ring radius is 539.99 $\mu m$. The bend loss of an ADS waveguide of this radius of curvature is negligible over the C-band; the mode is fully bound and only absorption and scattering contribute loss. Detailed data on absorption and scattering loss is not available beyond a specification of a total waveguide loss <0.5 dB cm$^{-1}$ with the expectation of a typical value of 0.1-0.2 dB cm$^{-1}$ for a ring of 50-100 $\mu m$ radii. The waveguide loss used in the simulation is 0.4 dB/cm as a conservative. A detailed simulation of the ring resonator can be found in reference [10].

Table 2. Design specifications detail of the AWG$_1$ and AWG$_2$.

| Design specifications | Comments | Remarks |
|---|---|---|
| AWG output channel spacing (GHz) | 50 | equal to the free spectral range of RR |
| AWG output channel bandwidth (GHz) | 20 | ~ ≤ 1/2 AWG output channel spacing |
| AWG input channel spacing (GHz) | 25 | half of AWG output channel spacing |
| Number of input channel | 8 | |
| Number of output channel | 32 | |
| Free spectral range FSR (GHz) | 1600 | 32×50 GHz |

Figure 4 (a-c) shows the measured add-drop transmission spectra of the RR at various wavelengths across the C-band. The peak transmission of the RR varies 1.25→ 1.75 dB from the longer wavelength edge to the shorter wavelength edge of the C- band. Fig. 4(d) shows the zoomed-in view of the ring resonance at the design wavelength. The full width half maximum (FWHM) bandwidth of the RR is found to be ~1.3 GHz. Figure 4(e-f) shows the measured FSR at the center wavelength and longer wavelength edge. Furthermore, the FSR is ~48.70 GHz at the edge of the short wavelength. The FSR variation is due to the effective index variation across the band from the index at the design wavelength, 1545 nm. This is due to the presence of group velocity dispersion. To minimise the relative detuning between the RR and AWG, the AWG channel passband center frequencies can be offset slightly. Further, digital signal processing can be used to provide the quantitative spectrum.

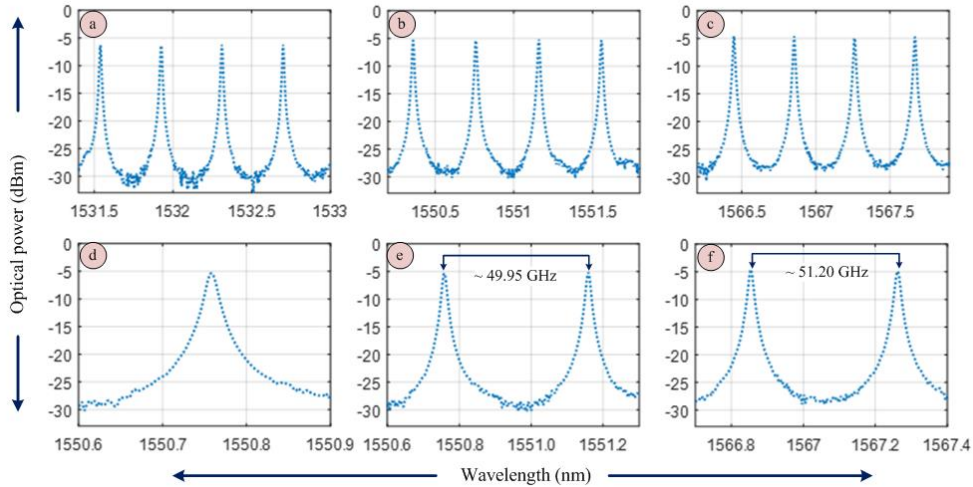

Fig. 4. Transmission spectrum of the RR obtained from fabricated chip using laboratory measurement; (a) smaller wavelength edge; (b) center wavelength; (c) longer wavelength edge; (d) zoom-in view of the ring resonance at the design wavelength. The measured full width half maximum (FWHM) is ~ 1.30 GHz; FSR at the (e) center wavelength and; (f) longer wavelength edge.

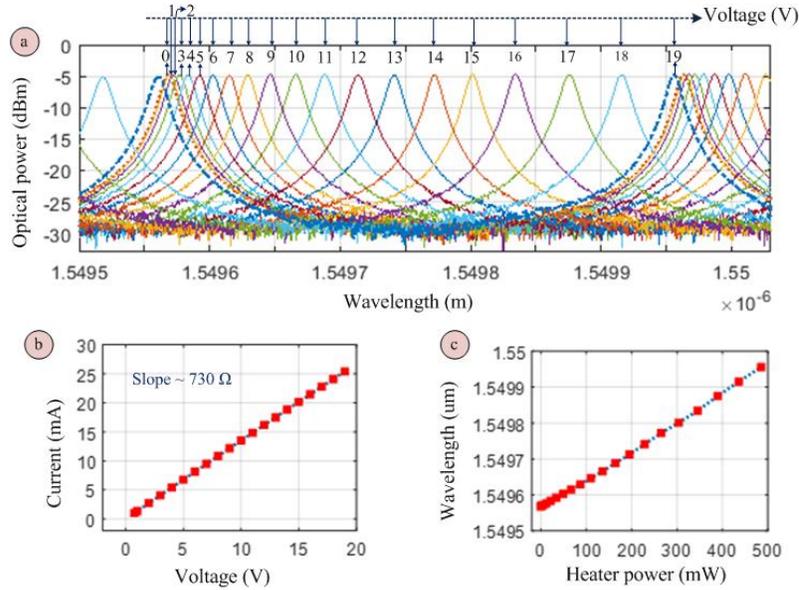

Figure 5. (a) Experimental demonstration of the tuning of the ring resonance as a function of applied voltage to the thermo-optic phase shifter; (b) The I-V characteristic of the thermo optic phase shifter and (c) Peak resonances in a particular FSR as a function of heater power.

The RR is tuned using an intra ring phase shifter. Only thermo-optic phase shifters are offered by the LioniX MPW process. Figure 5 (a) shows the tuning of the ring resonances as a function of applied voltage to the thermo-optic phase shifter. Tuning over a full FSR tuning is achieved. Figure 5 (b) shows the I-V characteristic of the heater while the frequency of the resonances as a function of heater power shown in Fig. 5(c). The phase shifter heater resistance is found to be 734 Ω by direct measurement which is well aligned with the slope (~730 Ω) of the I-V curve. The tuning wavelength (frequency) is found to be linearly proportional to the applied heater power as expected. However, due to low drive voltage, low drive power and linear voltage to index relationship, electro-optic tuning offers smooth operation with better efficiencies [22]. The use of piezo-electric actuators as an alternative means of providing an adjustable phase shift on the Si3N4 platform, augurs well for the future [21].

The long-arrayed waveguides of an AWG with 50 GHz channel spacing are vulnerable to process variations. In the case of an MPW fabrication run the resulting phase errors lead to process limited performance of 50 GHz; a custom fabrication run is necessary. Consequently, the measured ring resonator data is combined with AWG constructor data to simulated system operation. Two 32-channel cyclic AWG have been designed by Bright Photonics BV for fabrication to the specification given in Table 3. The AWG designs (AWG$_1$ & AWG$_2$) are identical apart from a 25 GHz relative shift of the channel spectra. Figure 6(a) shows the transmission spectra (truncated) of the upper AWG. Figure 6(b) shows a zoomed-in view of the interlaced AWG$_1$ and AWG$_2$ channel spectra when light is launched from input port 1. The aim of the reduction of the passband width from nominally 25 GHz to the 20 GHz specified is to reduce the intercept of adjacent channel spectra to the value of $\sim -22\ dB$ observed in Fig. 6(b). Similar results were obtained when light is launched from other input ports. For a ring resonator having FSR of 50 GHz, an AWG of 86~88 channels are required to cover the whole spectrum. The 32- channel AWG design presented here can easily be scaled up to 86 or 88 channels since a higher number (96) of output channels commercial AWG is already available [18]. Since the designed 32-channel AWG is cyclic, the output spectrum is periodic with a period of 32. Hence, with the aid of a tunable optical bandpass filter the spectrum of the whole C-band can be measured to prove the concept. Figure 6(c) shows the zoomed-out spectra of the AWG design. It shows that the envelop of the AWG passband is ~1.8 dB down at the edges of the AWG FSR compared to the center of the AWG FSR. This can be reduced significantly by increasing the FSR of the AWG.

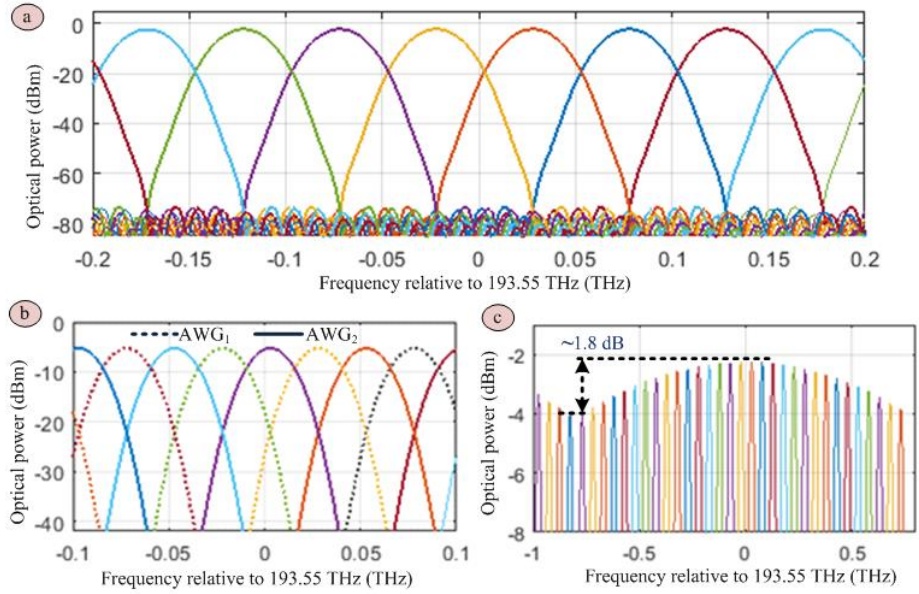

Fig. 6. (a) Simulated spectrum (truncated) of the AWG$_1$ with input light launch from upper left port (port 1); (b) zoom-in view of the interlaced spectrum between AWG$_1$ and AWG$_2$. The input light is launched from the identical input port (port 1); (c) Zoom-out transmission spectra of the AWG$_1$ output-channels.

The circuit level simulation has been implemented using VPIphotonics as shown in Fig. 7(a). The experimentally measured frequency response of the RR is imported in VPI simulation environment along with the characterization data of the $2 \times 2$ polarization maintaining splitter. The S matrix of the AWGs provided by Bright Photonics BV is imported into VPIphotonics. The scanning is performed by changing the position of the ring resonance across one FSR by applying voltage to the heater while recording the available optical power at the PIN diode at the output of each AWG channel. The weighted summation is performed offline to obtain the plots shown in Fig.7 (c). Figure 7(b) shows the comparison between the gaussian channel and constructor data. For the weighted summation method, gaussian channel weight is used. It shows that a ripple of ~1.70 dB is present in the detected power using constructor AWG channel response. The ripple can be reduced to ~0.80 dB using gaussian weighted summation as shown in Fig. 7(c). If one uses the frequency response obtained from constructor data as weight function, then the ripple will be zero as shown in Fig. 2(b). From the envelope of the AWG passband reported in Fig. (c), the measured power at the edges of the FSR will be ~1.5→2 dB less than the power at the center of the FSR.

The proposed weighted summation method can produce an identical response as like the center of the AWG FSR.

**Discussion**
For the simplicity of exposition, the peak of the ring resonance is aligned with the peak of the AWG channel spectra at zero intra-ring phase shift. This is not a necessary condition since the scanning is performed based on the simple weighted summation of the output of the AWGs. A wavelength meter [18] can be placed at an outer channel to monitor the position of the ring resonance with the channel. The only restriction of the design is that the frequency shift between $AWG_1$ and $AWG_2$ channels should be half (25 GHz in this case) of the AWG output channel spacing. If necessary, one of the AWGs can be temperature tuned to obtain the required frequency shift. In the ideal case (25 GHz shift), the measured optical power is symmetrical about the RR position. However, when the channels are slight detuned from the exact shift, the simulated output power breaks the symmetry. The first half of the RR tuning produces more ripple, whereas the ripple is reduced significantly in the second half of the RR and vice versa [11]. In the AWG design the adjacent channel cross talk is found to be greater than -25 dB. In practice, it may be well above -25 dB. The adjacent channel cross talk mainly depends on the phase errors in the arrayed waveguide sections. A ring resonator having FSR of 220 GHz using $Si_3N_4$ has already been reported in [3]. For an FSR of 200 GHz, only 22 AWG channels are required to cover the whole C-band. This leads to a compact design offering improved cross-talk performance. For example, an adjacent cross talk of $\sim -22\ dB$ is obtained for a $16 \times 16$ cyclic AWG using high contrast silicon photonics having 189 GHz output channel spacing [23]. On the other hand, an adjacent cross talk of $-18\ dB$ is obtained back in 1992 [24] for $15 \times 15$ multiplexer having 87 GHz output channel spacing using InP technology. A cross talk of $-13\ dB$ is obtained for 200 GHz output channel spacing using silicon nitride (SiN) technology [25]. Furthermore, AWG based on SiN platform having adjacent channel cross talk better than -39 dB with a channel spacing of 3.09 nm has already been reported in ref [26] long time ago. In practice, the cross talk can further be improved using high resolution optical lithography [27] along with a repeatable custom fabrication run. Finally, the simulation results obtained using components designed for fabrication bode well for sensing the whole C-band with high resolution bandwidth albeit with some ripple present in the detectable optical power which may be corrected, if required, by digital signal processing to provide the quantitative spectrum.

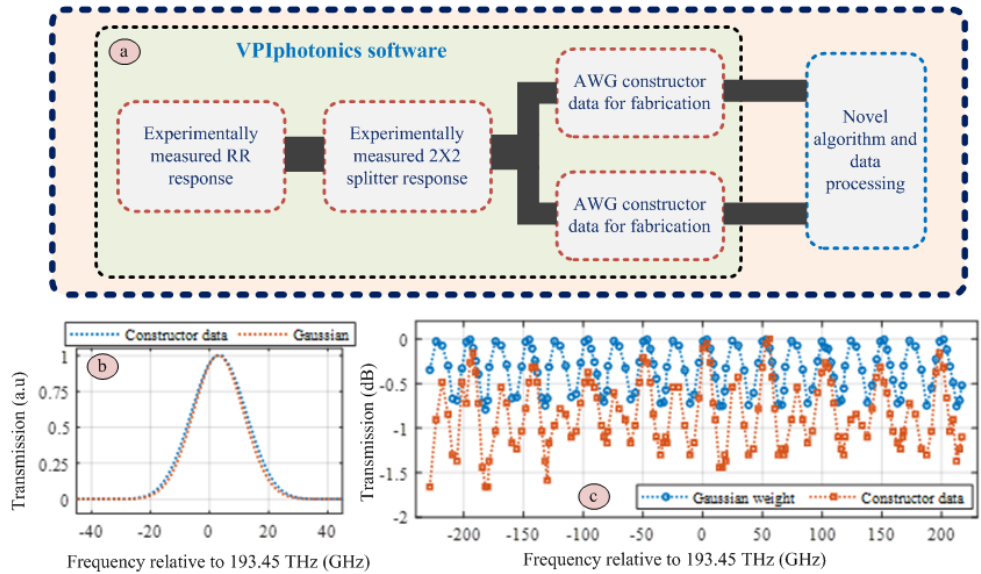

Fig. 7. (a) Schematic diagram of the simulation setup; (b) AWG channel frequency response comparison between gaussian channel and constructor data for fabrication; (c) overall result of the propose spectrometer using constructor data (without weight) and with gaussian weight implemented on the constructor data. Using gaussian weight reduces the ripple to ~0.8 dB from ~1.7 dB.

**Conclusion**
In summary, a simple circuit architecture for on-chip spectral monitor with high resolution is presented. The newly proposed signal processing method and feasibility for photonic integration places the spectrometer in the forefront of the state of the art. Detailed simulation results are presented using combination of experimental and constructor design data. As a main component of the proposed design, the integration feasibility of a high-resolution (~1.30 GHz) ring resonator is demonstrated experimentally. Tuning over a full FSR is achieved. The CMOS compatible $Si_3N_4$ platform is selected for fabrication due to its low loss and maturity. The incoherent summation architecture proposed herein is less compact but is more robust to fabrication process variations than the coherent summation architecture proposed previously. Based on the selection of proper weight, the spectral sensing can be done with minimum ripple to the flat detection for the best cases.

**Funding**
Huawei Canada sponsored research agreement 'Research on ultra-high resolution on-chip spectrometer'.

**Acknowledgments**
Mehedi Hasan acknowledges the Natural Sciences and Engineering Research Council of Canada (NSERC) for their support through the Vanier Canada Graduate Scholarship program. Trevor J. Hall is grateful to Huawei, Canada for their support of this work. Trevor J. Hall is also grateful to the University of Ottawa for their support of a University Research Chair.

**Disclosures**
The authors declare no conflicts of interest.